\documentclass[osajnl,twocolumn,showpacs,groupedaddress,10pt]{revtex4-1} 
\usepackage{amsmath,amssymb,graphicx}
\begin{document}

\title{Direct comparison of optical lattice clocks with an intercontinental baseline \\ of 9 000 km }

\author{Hidekazu Hachisu}\email{Corresponding author: hachisu@nict.go.jp}
\author{Miho Fujieda}
\author{Shigeo Nagano}
\author{Tadahiro Gotoh}
\author{Asahiko Nogami}
\author{Tetsuya Ido}
\affiliation{National Institute of Information and Communications Technology (NICT), 4-2-1 Nukui-Kitamachi, Koganei, Tokyo 184-8795, Japan}

\author{Stephan Falke}
\author{Nils Huntemann}
\author{Christian Grebing}
\author{Burghard Lipphardt}
\author{Christian Lisdat}
\author{Dirk Piester}
\affiliation{Physikalisch-Technische Bundesanstalt (PTB), Bundesallee 100, 38116 Braunschweig, Germany}

\begin{abstract}We have demonstrated a direct frequency comparison between two $^{87}{\rm Sr}$ lattice clocks operated in intercontinentally separated laboratories in real time. Two-way satellite time and frequency transfer technique based on the carrier phase was employed for a direct comparison with a baseline of 9 000 km between Japan and Germany. A frequency comparison was achieved for 83 640 s resulting in a fractional difference of $(1.1\pm1.6) \times 10^{-15}$, where the statistical part is the biggest contribution to the uncertainty. This measurement directly confirms the agreement of the two optical frequency standards on an intercontinental scale.
\end{abstract}


\maketitle 
Optical clocks have made rapid progress in the last ten years \cite{Poli}. Following an aluminum ion clock \cite{Al_NIST}, lattice clocks also reported an accuracy as well as an instability at the $10^{-18}$ level \cite{Sr_JILA}. Remote comparisons of optical clocks with this level of precision may work as a probe to detect differences in the gravitational redshift between different locations. Typical uncertainties in the level of the geoid surface currently amount to several centimeters, corresponding to differences in the redshift at the $10^{-17}$ level. Furthermore, with the high frequency stability of optical clocks, temporal variations of the gravitational potential due to tidal effects become relevant at the $10^{-17}$ level already at  distances of a few 100 km. This dynamical shift may be estimated by monitoring the clock comparisons on intercontinental scale because the differential tidal effects would become larger and observable in short time. With global frequency comparisons a variety of clock combinations may be established for testing fundamental postulates like a search for violations of Einstein's equivalence principle \cite{Clifford}. In metrological aspect, on the other hand, frequency agreement confirmed by intercontinental comparisons proves the capability of optical clocks to generate and maintain standard frequencies worldwide, which will support the optical redefinition of the second.

Frequency comparisons of optical clocks have been realized mostly on-campus \cite{Al_NIST,Sr_JILA,Yb_NIST,Taka,CaSr,Sr_SYRTE,NMIJ} except for two cases: one is a comparison of a $^{87}{\rm Sr}$ lattice clock at JILA against a neutral Ca clock at NIST with a 4-km-long optical fiber link \cite{SrCa_JILA} and the other is between two $^{87}{\rm Sr}$ lattice clocks at NICT and the University of Tokyo (UT) using a 60 km-long optical fiber \cite{UTLink}. Optical fiber links are promising on continental scales as demonstrated up to 1\nolinebreak 840 km \cite{PTBLinkPRL}. To bridge intercontinental distances, however, satellite based techniques are presently the only way. 
The global positioning system carrier-phase (GPSCP) technique, which has been the most precise satellite-based method, requires averaging times of more than a day to surpass an instability of $1 \times 10^{-15}$. As an alternative, carrier-phase based two-way satellite time and frequency transfer (TWCP), first demonstrated by U.S. Naval Observatory \cite{USNO}, is lately characterized in NICT for various ranges of the baselines up to 9 000 km which is realized between NICT and PTB \cite{MFIEEE, TWCPdetail}. Among satellite based techniques the TWCP technique has a superior short term instability (evaluated to be at the $10^{-13}$ level at 1 s) and is thus particularly suitable for comparing frequency standards with low instabilities, e.g. optical clocks.

In this Letter, two strontium lattice clocks, one at NICT in Japan \cite{Sr_NICT} and the other at PTB \cite{Sr_PTB, Sr_PTB2} in Germany, are directly compared using the TWCP technique. The agreement of these two optical clocks based on the same reference transition was confirmed with an uncertainty of $1.6\times 10^{-15}$. The baseline of 9 000 km is the longest in direct comparisons of optical clocks.

\begin{figure}[bthp]
\centerline{\includegraphics[width=.8\columnwidth]{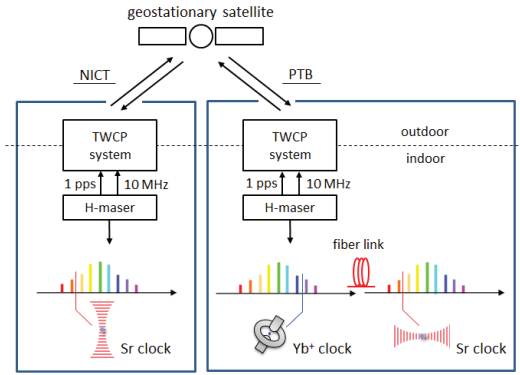}}
\caption{Schematic diagram of the frequency comparison of two $^{87}{\rm Sr}$ lattice clocks at NICT and PTB. Two-way satellite-based comparison using carrier phase measures the frequency difference of two local H-masers, via which the differential frequency of two clocks is derived. pps: pulse per second, TWCP: two-way carrier phase (time and frequency transfer) }
\end{figure}

Figure 1 shows the setup of the comparison. It consists of a TWCP setup, $^{87}{\rm Sr}$ lattice clocks and frequency combs at each site. Both lattice clocks use the transition $^1S_0(F=9/2) - {}^3P_0(F=9/2)$ as reference. The accuracy of the $^{87}{\rm Sr}$ lattice clock at NICT and PTB is $2 \times 10^{-16}$ and $4 \times 10^{-17}$, respectively. Frequencies of the lattice clocks at each site are measured referenced to a local hydrogen maser (H-maser). At NICT, a Ti-sapphire based frequency comb is stabilized to the H-maser and the clock frequency $\nu_{\rm NICT}$ is obtained from the beat signal against the frequency comb. At PTB, the transfer oscillator method \cite{transferO} is used to measure frequency ratios with fs-combs. The H-maser and the Sr clock are located in different buildings each with an erbium-fiber based frequency comb. These frequency combs are connected by stabilized optical fiber links. The frequency of the  lattice clock at PTB $\nu_{\rm PTB}$ is derived from relevant beat signals with a reference to the local H-maser. Additionally an $^{171}{\rm Yb}^+$ ion clock based on the electric octupole transition $^2S_{1/2}(F=0) - {}^2F_{7/2}(F=3)$ is operated at PTB with a fractional uncertainty of less than $7 \times 10^{-17}$ \cite{YbIon_PTB, YbIon_PTB2}. Both the lattice clock and ${\rm Yb}^+$ clock are stable enough to assume that the frequency ratio of two clock frequencies is constant with an uncertainty at the  $10^{-17}$ level\cite{Sr_PTB2}. Thus, the ${\rm Yb}^+$ clock can extend the observation time when the PTB lattice clock is off-line. Note that zero dead-time counters are used in all part of the experiment.

In terms of the TWCP setup, we use a frequency band on a transponder of the geostationary satellite AM2, which is located at longitude $80^\circ$ East. The transponder is available from 10:05 to 22:59 UTC per day. The elevation angles at NICT and PTB are $16.0^\circ$ and $3.7^\circ$. In the two-way method both ground stations transmit and receive microwave signals simultaneously. The carrier frequencies of the uplink and the downlink are about 14 GHz and 11 GHz. The uplink signals are referenced to the local H-masers while the downlink signals are measured against these H-masers. The carriers are modulated by narrow-band pseudo random noise to avoid fading and other perturbations. The group delay and the carrier phase of the received signals were detected by demodulation using a replica code, by which we derived the fractional frequency difference of the two H-masers. The technical details of the TWCP system are described in \cite{MFIEEE, TWCPdetail}. The frequency ratio of the $^{87}{\rm Sr}$ lattice clocks is obtained through the ratio of the two H-masers in real time.

Figure 2 shows a typical time record of the fractional difference $\Delta \left( t\right) =\nu_{\rm NICT}( t ) / \nu_{\rm PTB} (t) -1$. Each data point is the average of 60 s signal integration. The fractional difference includes corrections due to the systematic shifts of the clocks  as well as the gravitational red shift of $2.4(1.0) \times 10^{-16}$\cite{Sr_NICT, PTBCsF2, tidal}. The frequency difference between uplink and downlink causes different delays due to the dispersion of the ionosphere. A compensation for this effect is applied as discussed in \cite{TWCPdetail}. We checked the data of Fig. 2 as well as the whole dataset for its sample autocorrelation  $\rho \left( \tau_{\rm lag} \right)= \overline{\Delta\left( t + \tau_{\rm lag}\right)\Delta \left( t\right) }/ \sigma_\Delta^2$ at lag $\tau_{\rm lag}$, where $\sigma_\Delta^2$ is the variance of $\Delta \left( t\right)$ \cite{Zhang}. It is larger than 0.2 for the range of $0< \tau_{\rm lag} <150$ s and stays above 0.1 up to 700 s. This suggests that the clock comparison is neither characterized by white phase noise nor by white frequency noise.

\begin{figure}[bthp]
\centerline{\includegraphics[width=.9\columnwidth]{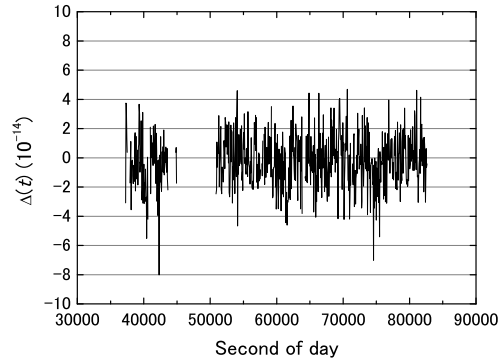}}
\caption{Record of the fractional frequency difference of one typical experimental day. Corrections of systematic shifts of the atomic clocks as well as the compensation of ionospheric delays are included. Each point is the result of 60 s signal integration. }
\end{figure}

The Allan deviation of $\Delta\left( t\right)$ of the longest continuous recording (30 900 s shown as the later part of Fig. 2) is presented as the filled circles in Fig. 3. The reduction of the Allan deviation with the signal integration time is slower than in case of white frequency noise, which confirms the autocorrelation analysis described above. We fitted the instability obtaining $7.5 \times 10^{-14} \, \tau ^{-0.37}$. Possible sources of instabilities are ambient temperature changes affecting the outdoor microwave equipment or imperfections of the ionosphere corrections.

\begin{figure}[hthp]
\centerline{\includegraphics[width= .9\columnwidth]{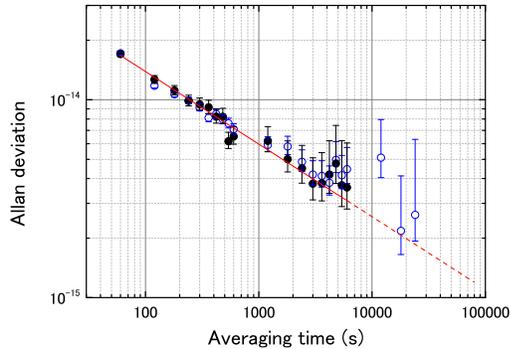}}
\caption{Allan deviation of the frequency ratio versus the averaging time. The instabilities shown as filled circles is derived from the later part of the Fig. 2, which contains 30 900 s of continuous data. Red solid line and red dashed line show the weighted fit and its extrapolation. Open circles are obtained by treating the whole 83 640 s as one continuous measurement in which a part of 13 840 s relies on the $\rm{Yb}^+$ clock. Instability is gradually reduced with a slope of $\tau^{-0.37}$. }
\end{figure}
We operated the clocks for several hours per day over four days and obtained simultaneous operation of the two lattice clocks for 69 840 s in total.  This data set was extended by 13 800 s using the ${\rm Yb}^+$ ion clock at PTB, resulting in a total measurement of 83 640 s. The conversion coefficient of ${\rm Yb}^+$ flywheel oscillator was derived from the measurement of 45 960 s, in which two optical clocks at PTB were both operated during the measurement campaign. The Allan deviation of this extended data set is shown in Fig. 3 as open circles. It agrees with the long term instability obtained from extrapolation of the longest continuous data set presented as filled circles.


The observed instability results from link instabilities as the optical clocks have demonstrated much smaller instabilities. An upper limit for the link instabilities is obtained from the comparison of the TWCP link
against the GPSCP link which is totally independent from the TWCP. The comparison was done by measuring UTC(NICT) versus UTC(PTB) via the two link techniques during the four days of the measurement discussed here as well as in a preceding study over two months. The Allan deviations of the link comparisons are shown in Fig. 4, which provides an upper limit of the instability of each of the two satellite link techniques. As the measurement time of the optical clock comparison was limited, its instability did not fall below $10^{-15}$ (Fig. 3) but is below the combined instability of the GPSCP and the TWCP links (Fig. 4). We estimate the statistical uncertainty of the frequency comparison by an extrapolation to the total measurement time using the observed instability progression and find $1.2\times 10^{-15}$ for 83 640 s measurement time. This is consistent with the comparison of the two link techniques. The averaged frequency difference between the two link techniques was less than $1.0\times 10^{-15}$ \cite{TWCPdetail}, which we assign as the systematic uncertainty of the link for the clock comparison. The impact of intermittent availability of the link is already included when we use this averaged frequency difference 
as the systematic uncertainty.

\begin{figure}[hthp]
\centerline{\includegraphics[width=0.9 \columnwidth]{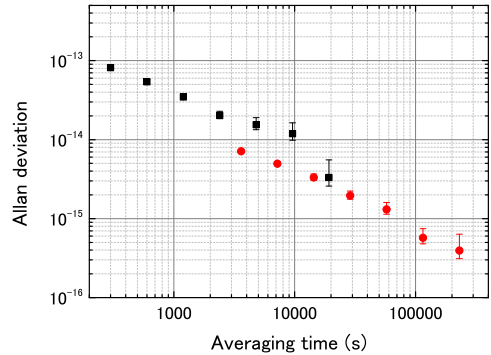}}
\caption{Combined instability of satellite links: Allan deviation of the difference between the frequency comparisons of UTC(NICT) and UTC(PTB) obtained via TWCP and GPSCP.
Filled squares in black show the performance during the four days as the direct frequency comparison of two $^{87}{\rm Sr}$ clocks while filled circles in red represent the measurement of March and April in 2013.}
\end{figure}

By combining the systematic uncertainties of the lattice clocks ($2 \times10^{-16}$ and $4 \times 10^{-17}$) and the link ($1.0 \times 10^{-15}$) as well as the statistical uncertainty ($1.2 \times 10^{-15}$) we obtain the fractional uncertainty of the clock comparison to be $1.6 \times 10^{-15}$. Using the fractional frequency difference $\Delta =1.1 \times 10^{-15}$ along with a recent absolute frequency measurement of the PTB strontium lattice clock with an uncertainty of $3.9 \times 10^{-16}$ \cite{Sr_PTB2}, the absolute frequency of the NICT strontium lattice clock is derived to 429 228 004 229 873.60 (71) Hz.

Figure 5 relates our clock comparison to other methods of remote frequency standard evaluations: An example for a precise comparison via local realizations of the second with Cs fountain clocks is given by the comparison of the Sr lattice clocks of PTB and SYRTE \cite{Sr_PTB2, Sr_SYRTE} ($6 \times 10^{-16}$ uncertainty). In the absence of a local primary clock in operation, TAI corrected H-masers are used as reference and the obtained frequency can be compared to published frequency values, e.g. a recent measurement of NICT \cite{Sr_NICT} versus the weighted mean of ten published strontium frequencies 
, resulting in a moderate agreement of $1.8(3.3) \times 10^{-15}$. Moreover, one introduces a latency of a month to obtain the correction of the local UTC realization. Currently the most accurate method for frequency comparisons are stabilized optical fiber links, through which optical clocks can be compared without significant loss of accuracy or stability \cite{PTBLinkPRL}. The third data point of Fig. 5  taken from the NICT vs. UT comparison \cite{UTLink} represents such a measurement with an uncertainty of $7 \times 10^{-16}$ that is mostly due to the compared clocks. The result presented in this letter (bottom point in Fig. 5) has slightly larger uncertainty, but it is provided in real time and was obtained requiring neither a primary standard nor a fiber link.

\begin{figure}[hthp]
\centerline{\includegraphics[width=\columnwidth]{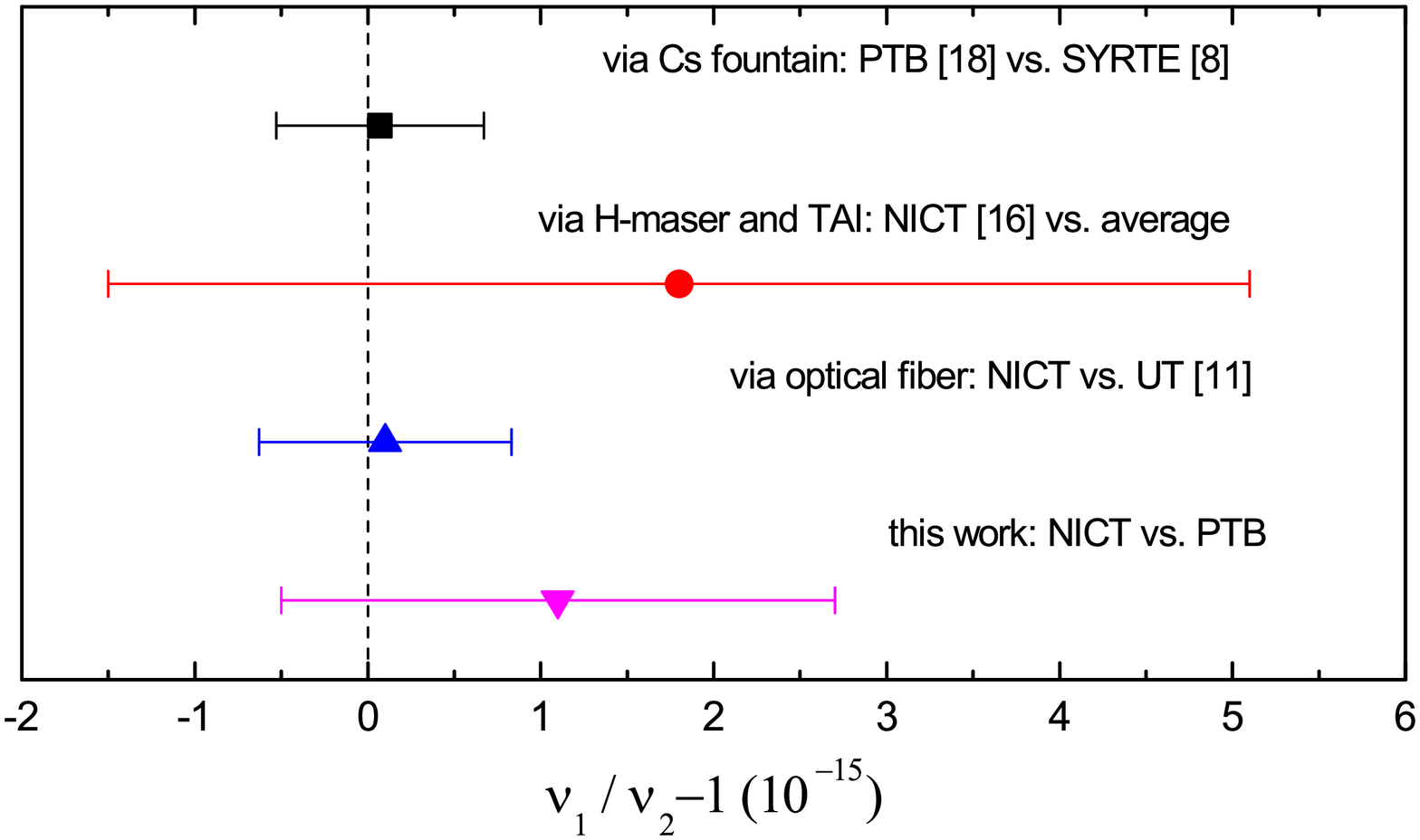}}
\caption{Overview of frequency comparisons of remote strontium lattice clocks using different methods.}
\end{figure}

In summary, a direct frequency comparison between intercontinentally distant optical clocks was for the first time demonstrated using the TWCP method. Two strontium lattice clocks are located in NICT (Japan) and PTB (Germany) with the baseline of 9 000 km. The clock frequencies agree with $1.1(1.6) \times 10^{-15}$, dominated by the uncertainty contribution of the link. Further improvements may be achieved with more continuous measurement time and more careful control of ambient temperature of the link system.  We expect to reach the $10^{-16}$ level of accuracy in the near future with commercially available ground stations modified to apply the TWCP technique. Optically generated microwaves \cite{OPMW1, OPMW2} may replace H-masers as local reference to synthesize the carrier frequency. Baselines capable of connecting continents and transportable ground stations may further fuel the interest in the demonstrated technique.

We thank N. Shiga, K. Kido, H. Ito, Y. Hanado, A. Bauch, J. Becker, and E. Peik for useful discussions and technical support. This research was supported in part by the FIRST Program
of the Japan Society for the Promotion of Science, the Centre of Quantum Engineering and Space-Time Research (QUEST), and the projects ‘International Timescales with Optical Clocks’ (ITOC) and ‘High-accuracy optical clocks with trapped ions’, which are both part of the European Metrology Research Programme (EMRP). The EMRP is jointly funded by the EMRP participating countries within EURAMET and the European Union.


\begin{thebibliography}{99}
\bibitem{tidal} Temporal tidal effect is conservatively considered.
\end{thebibliography}
\end{document}